\documentclass[twocolumn,jkps,fleqn,showpacs,showkeys,floatfix]{revtex4}
\usepackage[pdftex]{graphicx}
\usepackage{amssymb}
\usepackage{amsmath}
\usepackage{bm}
\usepackage{kotex}

\usepackage{xcolor}

\usepackage{amsmath,amssymb,mathtools}
\usepackage{bm}
\usepackage{booktabs}

\begin{document}
\setcounter{page}{1}
\title[]{Linear Temporal Structure and Short-Term Persistence of Conflict Activity in Middle Eastern Countries}
\author{Hyuntae \surname{Ahn}}
\author{Mi Jin \surname{Lee}}
\email{mijinlee@pusan.ac.kr}
\affiliation{Department of Physics, Pusan National University, Busan 46241, Republic of Korea}


\begin{abstract}
Conflict events in Middle Eastern countries exhibit complex temporal fluctuations. As a simple baseline, we examine whether their conflict-frequency time series contain measurable linear temporal structure using the autoregressive integrated moving average (ARIMA) framework, which assumes linear dependence on past observations and errors. Despite this simple linear assumption, the temporal dependence is largely captured for most countries by ARIMA. Predictions based on the selected ARIMA models reproduce the overall temporal behavior to some extent, although abrupt changes remain difficult to predict. We further examine the short-term persistence of changes in conflict activity using the first zero-crossing point $k^*$, which characterizes the duration of the initial positive temporal correlation, and the integrated correlation time $\tau_{\mathrm{int}}$, which additionally reflects its strength and decay. The resulting $k^*$ values are approximately 2--6 days, while $\tau_{\mathrm{int}}$ ranges from approximately 0.6 to 1.7 days. The significant negative association of $\tau_{\mathrm{int}}$ with mean GDP per capita suggests a possible relation between short-term conflict persistence and broader socioeconomic conditions. These results reveal measurable linear temporal structure and country-dependent short-term persistence in the temporal patterns of conflict activity across the Middle East.
\end{abstract}
\keywords{temporal dependence; conflict events; ARIMA; persistence}

\maketitle

\section{Introduction}
\label{sec:introduction}
Conflict activity changes irregularly over time under the influence of complex historical, political, and social factors. Empirical studies have reported temporal dependence in conflict dynamics. Examples include temporal dependence in conflict events during the Bosnian war~\cite{Weidmann2010predicting}, positive temporal autocorrelation in daily fatalities during the Syrian conflict~\cite{Fujita2017correlations}, persistent increases in event occurrence following violent incidents in Iraq described by self-exciting point-process models~\cite{Lewis2012Iraq}, and spatiotemporally connected chains of conflict events~\cite{Kushwaha2023Mesoscale}. Such temporal information has also been exploited for prediction: early-warning systems for political violence have been developed~\cite{Hegre2019ViEWS}, while more recent studies have identified predictive information in past conflict levels and temporal patterns~\cite{chadefaux2025endogenous,Schincariol2025variability}. These findings suggest that conflict dynamics can retain information from their recent past despite highly irregular fluctuations. Quantifying such temporal dependence may therefore help characterize the dynamics and the predictability of conflict activity.

The Armed Conflict Location \& Event Data (ACLED)~\cite{ACLED2025Data,Raleigh2023ACLED} provide detailed event-level information on individual conflict events, including their occurrence dates, locations, actors, and event types. Such event-based data have also been used to compare conflict dynamics across countries and regions~\cite{Browning2026Hawkes}. Here, we focus on Middle Eastern countries, where conflict events have been persistently recorded across multiple countries during the study period, making the region suitable for comparing country-level temporal patterns of conflict activity. We aggregate conflict events over fixed time intervals to construct a conflict-frequency time series for each country. Although this representation simplifies the detailed context of individual events, it provides a common observable for comparing the macroscopic temporal variation of conflict activity across countries.

As a baseline, we employ the autoregressive integrated moving average (ARIMA) framework~\cite{Box2015TimeSeries} to examine the linear temporal structure of these conflict-frequency time series. ARIMA and related models have previously been applied to conflict- and protest-event time series primarily for modeling and forecasting~\cite{RodriguezWhite2023CrimeScience,Browning2026Hawkes}. These applications mainly focus on constructing and validating predictive models. Here, we instead use ARIMA as a diagnostic baseline to ask how much of the observed temporal variation can be described by linear combinations of past observations and past errors. Despite this simple linear assumption, our results show that substantial linear temporal structure is present across most Middle Eastern countries. The resulting predictions also reproduce the overall temporal behavior to some extent, although their accuracy varies across countries and abrupt changes remain difficult to reproduce.

The linear temporal structure, however, does not directly provide a characteristic time scale over which changes in conflict activity persist. Previous studies have examined temporal dependence in conflict occurrence, fatalities, and sequences of individual events~\cite{Weidmann2010predicting,Fujita2017correlations,Lewis2012Iraq,Kushwaha2023Mesoscale}, whereas the persistence of increases and decreases in conflict activity itself has received relatively less attention. We therefore examine the autocorrelation of local changes in conflict frequency and characterize their short-term persistence using the first zero-crossing time and the integrated correlation time. The former measures the duration of the initial positive temporal correlation, while the latter additionally reflects its strength and decay. We find that the initial positive temporal correlation persists only over a few days, with the corresponding time scales varying across countries. We further examine the relation between the integrated correlation time and country-level socioeconomic conditions and find a significant negative association with mean GDP per capita, suggesting a possible connection between the persistence of conflict-activity changes and broader socioeconomic conditions.

The rest of this paper is organized as follows. The conflict-event data are described in Sec.~\ref{sec:data}. Section~\ref{sec:arima} introduces the ARIMA model and presents the corresponding analysis. The short-term correlation time is analyzed in Sec.~\ref{sec:tau}. Finally, Sec.~\ref{sec:discussion} provides further discussion.

\section{Data and Construction of Conflict-Frequency Time Series}
\label{sec:data}
\begin{figure}
    \centering
    \includegraphics[width=\columnwidth]{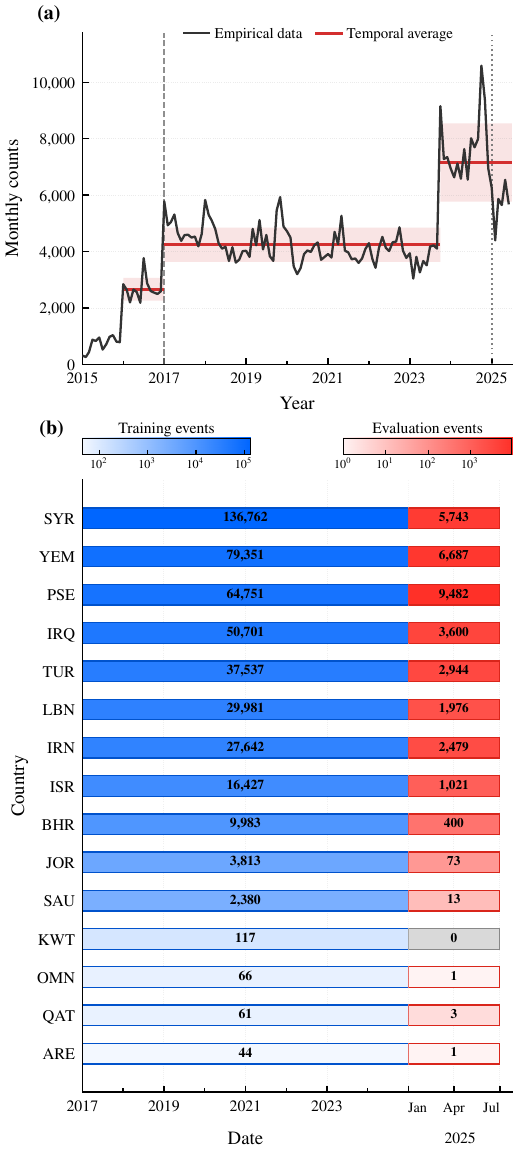}
    \caption{Time series of conflict events. (a) Monthly aggregated time series over 15 Middle Eastern countries from January 2015 to June 2025. The black line represents the empirical data. The red solid lines and shaded areas indicate the temporal averages and standard deviations, respectively, over the corresponding periods. (b) Training and evaluation periods for each country. The left and right bars indicate the training and evaluation periods, respectively, while the colors represent the corresponding numbers of conflict events. KWT is excluded from the ARIMA evaluation because no conflict events are recorded during the 2025 evaluation period.}
    \label{fig:data_statistics}
\end{figure}

We analyze time series of conflict events in 15 Middle Eastern countries using data curated by the Armed Conflict Location \& Event Data (ACLED)~\cite{Raleigh2023ACLED,ACLED2025Data}. We include all event records contained in the ACLED dataset used in this study. The countries are ARE, BHR, IRN, IRQ, ISR, JOR, KWT, LBN, OMN, PSE, QAT, SAU, SYR, TUR, and YEM, with their full names listed in Table~\ref{tab:code} in Appendix~\ref{app:code}. The dataset contains a total of $536\,433$ conflict events recorded from January 1, 2015 to July 11, 2025. Each record corresponds to an individual conflict event and includes 31 types of metadata, such as the event date, event type, and number of fatalities. We use the event date recorded for each ACLED event to construct the temporal series. For visualization, the monthly aggregated time series over all 15 countries is shown in Fig.~\ref{fig:data_statistics}(a).

Conflict activity across the Middle East shows distinct temporal patterns, including abrupt jumps (e.g., in 2016, 2017, and late 2023) and deviations from the average behavior. In particular, the temporal average behavior remains relatively stable from 2017 until late 2023. Based on these observations, we consider the period from January 1, 2017 to December 31, 2024, which contains a total of $459\,616$ events across the 15 countries. The subsequent period from January 1 to June 30, 2025 contains a total of $34\,423$ events and is used for model evaluation. For the time-series construction, intervals with no recorded conflict events within the common analysis period are assigned a value of zero. The time-series data for the individual countries are summarized in Fig.~\ref{fig:data_statistics}(b), where only nonzero event records are displayed for visualization. KWT is excluded from the 2025 evaluation because no conflict events are recorded during this period, leading to 14 countries for analysis.

Daily aggregation provides a large number of data points but exhibits strong day-to-day fluctuations, whereas monthly aggregation produces a smoother time series at the cost of substantially fewer data points. We also tested the ARIMA analysis with $\Delta T=1$~day, but the model did not perform adequately for the daily series. We therefore adopt $\Delta T=1$~week, which reduces short-term fluctuations while retaining sufficient temporal resolution, and construct a weekly aggregated time series for each country. We define the daily conflict count in country $c$ on day $t$ as $n_t^{(c)}$. The weekly aggregated conflict count is then given by
\begin{equation}
    x_t^{(c)}=\sum_{\tau=t}^{t+\Delta T-1} n_\tau^{(c)},
\label{eq:xt}
\end{equation}
where $t=t_0,t_0+\Delta T,t_0+2\Delta T,\cdots$. Thus, $x_t^{(c)}$ represents the number of events that initially occur in country $c$ within the non-overlapping interval $[t,t+\Delta T)$. 

\section{Linear Temporal Structure and Prediction with ARIMA}
\label{sec:arima}
\subsection{Model and selection procedure}
To quantify the linear temporal dependence described above, we employ the autoregressive integrated moving average (ARIMA) model~\cite{Box2015TimeSeries}, which represents the current state in terms of past observations and past errors. ARIMA requires the time series to be stationary after appropriate differencing. When the original series is non-stationary, differencing is performed until stationarity is achieved, with the number of differencing operations defining the order $d$.

Stationarity is assessed using the augmented Dickey--Fuller (ADF)~\cite{Said1984ADF,Dickey1979DF} and Kwiatkowski--Phillips--Schmidt--Shin (KPSS)~\cite{Kwiatkowski1992KPSS} tests. The ADF test examines whether the effect of a shock persists over time, whereas the KPSS test examines whether fluctuations remain stable around the overall level of the series. Because the two tests assess stationarity from complementary viewpoints, we use them together to determine whether differencing is required. Depending on the stationarity of $x_t$ in Eq.~(\ref{eq:xt}), differencing of order $d$ is applied: e.g., $x_t$ for $d=0$, $x'_t=x_t-x_{t-\Delta T}$ for $d=1$, and $x''_t=x'_t-x'_{t-\Delta T}$ for $d=2$.

Let $y_t$ denote the stationary series obtained after differencing. The ARIMA model assumes
\begin{equation}
y_t = \mu_d + \sum_{i=1}^p \phi_i y_{t-i} + \sum_{j=1}^q \theta_j \epsilon_{t-j} + \epsilon_t,
\label{eq:ARIMA_assume}
\end{equation}
where $\mu_d$ denotes the constant term, with $\mu_d=0$ for $d=1$ in the present implementation, $\phi_i$ and $\theta_j$ weight past observations and errors, respectively, and $\epsilon_t$ is a random error. The orders $p$ and $q$ specify how many past observations and past errors are taken into account, respectively and are measured in units of $\Delta T=1$~week. For a given set of $(p, d, q)$, the coefficients $\mu_d$, $\phi_i$, and $\theta_j$ are estimated from the training data using maximum likelihood estimation. The resulting model is denoted by ARIMA$(p,d,q)$. For example, ARIMA$(0,0,0)$ and ARIMA$(0,1,0)$ correspond to white noise and a random walk, respectively.

Let $s$ denote the forecast origin and $h$ the forecast horizon. The $h$-step-ahead forecast is given by
\begin{equation}
\hat{y}_{s+h|s}=\mu_d+\sum_{i=1}^p\phi_i\tilde{y}_{s+h-i|s}+\sum_{j=1}^q\theta_j\tilde{\epsilon}_{s+h-j|s},
\label{eq:ARIMA_predict}
\end{equation}
where
\begin{equation}
\tilde{y}_{r|s}=\begin{cases}y_r,&r\leq s,\\ \hat{y}_{r|s},&r>s,\end{cases}\qquad \tilde{\epsilon}_{r|s}=\begin{cases}\hat{\epsilon}_{r|r-1},&r\leq s,\\ 0,&r>s.\end{cases}
\label{eq:ARIMA_recursive}
\end{equation}
The error with empirical $y_t$ is defined as
\begin{equation}
\hat{\epsilon}_{t|s}=y_t-\hat{y}_{t|s}.
\label{eq:residual}
\end{equation}
Within the training period, setting $s=t-1$ and $h=1$ gives the one-step-ahead fitted value $\hat{y}_{t|t-1}$ and the corresponding residual $\hat{\epsilon}_{t|t-1}$. For prediction beyond the training period, we set $s=T$, where $T$ is the last training time, and evaluate $\hat{y}_{T+h|T}$ recursively for $h=1,2,\cdots$.

Then, how do we determine the optimal ARIMA($p,d,q$)? For a given $d$, the optimal $p$ and $q$ are selected using the Akaike information criterion (AIC)~\cite{Akaike1974AIC}, Bayesian information criterion (BIC)~\cite{Schwarz1978BIC}, and Hannan--Quinn information criterion (HQIC)~\cite{Hannan1979HQIC}. These criteria balance model fit and model complexity, with AIC placing relatively more emphasis on fit, BIC imposing a stronger complexity penalty, and HQIC providing an intermediate penalty (see Appendix~\ref{app:ARIMA_diagnostics}). The optimal $p$ and $q$ are determined by minimizing each criterion (see Fig.~\ref{fig:arima_grid}).

If the linear ARIMA model adequately captures the temporal dependence in the data, no systematic structure should remain in the residuals $\hat{\epsilon}_{t|t-1}$~\cite{Ljung1978Box}. We verify this using a statistical test for residual autocorrelation (see Appendix~\ref{app:ARIMA_diagnostics}).

\subsection{Selection and adequacy of ARIMA model}
\begin{table}[t]
\centering
\caption{The optimal parameters for the ARIMA model. The differencing order $d$ is determined by the ADF and KPSS tests. The orders $p$ and $q$ of past observations and errors are obtained by minimizing the three information criteria AIC, BIC, and HQIC. Models that are not adequate according to the statistical test (see details in Appendix.~\ref{app:ARIMA_diagnostics}) are marked by cross symbols.}
\label{tab:arima_models}
\setlength{\tabcolsep}{4.2pt}
\begin{tabular}{c|cccc}
\hline
Country & $d$ & AIC $(p,q)$ & BIC $(p,q)$ & HQIC $(p,q)$ \\
\hline
ISR & 1 & $(2,8)$  & $(1,1)\,\times$ & $(1,1)\,\times$ \\
SAU & 1 & $(4,3)$  & $(1,2)\,\times$ & $(6,0)$ \\
IRQ & 1 & $(3,4)$  & $(1,1)$ & $(3,4)$ \\
BHR & 1 & $(1,6)$  & $(1,1)\,\times$ & $(0,6)$ \\
TUR & 1 & $(1,8)$  & $(1,1)\,\times$ & $(3,3)$ \\
SYR & 1 & $(1,1)$  & $(1,1)$ & $(1,1)$ \\
PSE & 1 & $(1,2)$  & $(1,2)$ & $(1,2)$ \\
JOR & 0 & $(2,1)$  & $(2,1)$ & $(2,1)$ \\
IRN & 1 & $(1,2)$  & $(1,2)$ & $(1,2)$ \\
LBN & 1 & $(10,1)$ & $(1,1)\,\times$ & $(1,1)\,\times$ \\
YEM & 1 & $(1,1)$  & $(1,1)$ & $(1,1)$ \\
QAT & 0 & $(1,0)$  & $(1,0)$ & $(1,0)$ \\
OMN & 0 & $(0,1)$  & $(0,1)$ & $(0,1)$ \\
ARE & 0 & $(1,0)$  & $(1,0)$ & $(1,0)$ \\
\hline
\end{tabular}
\end{table}

\begin{figure*}[!t]
\centering \includegraphics[width=\linewidth] {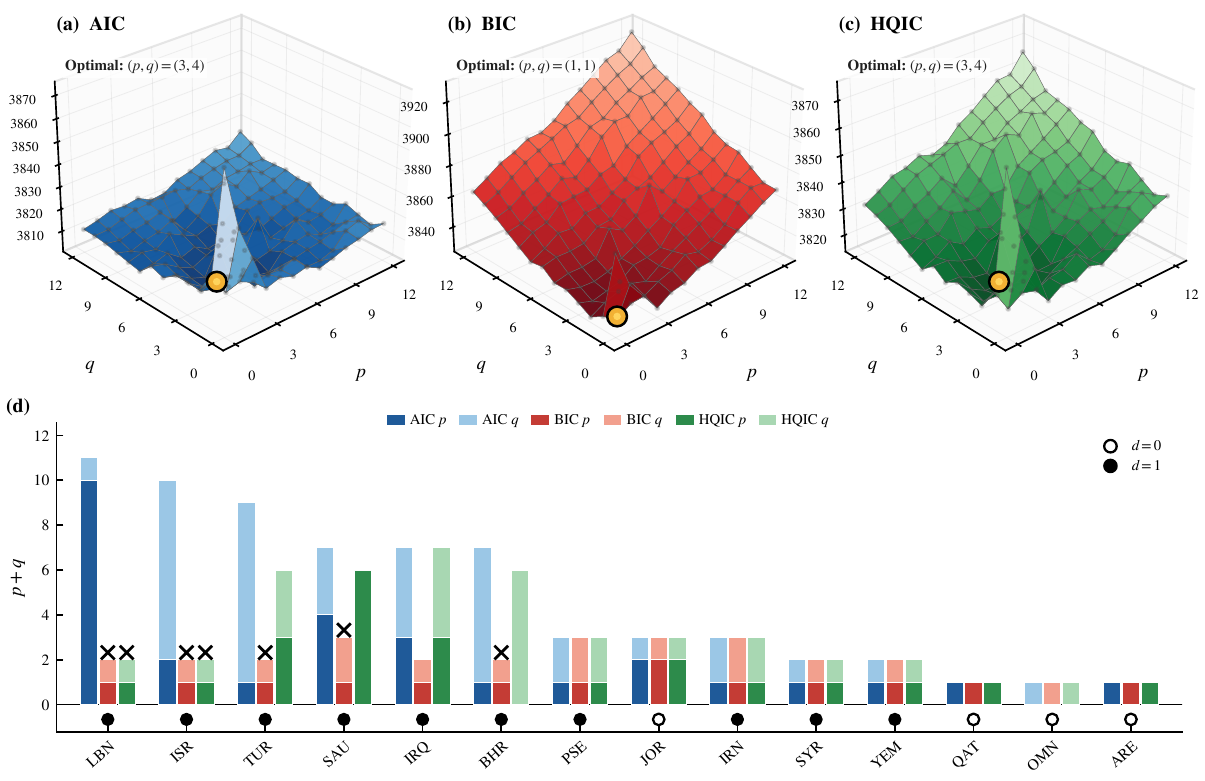} \caption{Optimal parameters $(p,q)$ for a given $d$ for IRQ obtained using (a) AIC, (b) BIC, and (c) HQIC. The colors correspond to the values of the information criteria, and the yellow circles indicate the optimal points. (d) Results for all 14 countries except KWT. The bar height corresponds to $p+q$, and the lower and upper boxes represent $p$ and $q$, respectively. The open and filled circles indicate $d=0$ and $d=1$, respectively. The numbers of $(p, q)$ are listed in Table~\ref{tab:arima_models}. Parameter sets that are not adequate according to the statistical test (see details in Appendix.~\ref{app:ARIMA_diagnostics}) are marked by cross symbols.}
\label{fig:arima_grid}
\end{figure*}

Applying the procedure described above to the weekly conflict-count series from 2017 to 2024, we determine the ARIMA orders $(p,d,q)$ for each country, using Eqs.~(\ref{eq:ARIMA_predict})--(\ref{eq:residual}) and the relevent statistical tests described in Sec.~\ref{sec:arima}. Four countries (JOR, QAT, OMN, and ARE) show stationary behavior in the original series ($d=0$), whereas the remaining countries require first-order differencing ($d=1$), as shown in Table~\ref{tab:arima_models}. Accordingly, $y_t=x_t$ is used for the former four countries, whereas $y_t=x'_t$ is used for the remaining countries.

For each stationary series $y_t$, we evaluate AIC, BIC, and HQIC over the $p$-$q$ plane and identify the minimum of each criterion. The results for IRQ are illustrated in Figs.~\ref{fig:arima_grid}(a)--\ref{fig:arima_grid}(c), where AIC and HQIC select $(p,q)=(3,4)$, while BIC selects $(p,q)=(1,1)$. The selected $(p,q)$ values for all countries are listed in Table~\ref{tab:arima_models} and visualized in Fig.~\ref{fig:arima_grid}(d). For eight of the fourteen countries, AIC, BIC, and HQIC select the same pair of $(p,q)$, indicating that the selected parameter set is robust to the choice of information criterion. For the remaining countries, including IRQ, the selected orders vary across criteria, indicating a stronger dependence of the parameter selection on how model complexity is penalized.

To evaluate the adequacy of the selected ARIMA parameter sets, we examine the one-step-ahead residuals defined in Eq.~(\ref{eq:residual}). The residual autocorrelations for ARIMA$(3,1,4)$ selected by AIC for IRQ are statistically insignificant (see Fig.~\ref{fig:arima_residual}), indicating that no systematic temporal correlation remains and that this parameter set adequately captures the linear temporal dependence. The residual distribution is approximately normal, although some extreme fluctuations with large $|\hat{\epsilon}_{t|t-1}|$ deviate from the normal distribution. While normality is not required for the residuals to be temporally uncorrelated, this deviation indicates that the Gaussian approximation is less accurate for extreme fluctuations (see Appendix~\ref{app:ARIMA_diagnostics} for details).

Most of the other selected parameter sets also show no significant residual autocorrelation, except for a few cases as marked by $\times$ symbols in Fig.~\ref{fig:arima_grid}(d) and Table~\ref{tab:arima_models}. These few exceptions, excluded from subsequent analysis, indicate that some linear temporal dependence remains unexplained for the corresponding parameter sets.

\subsection{Evaluation results for 2025}
\begin{figure}
    \centering
    \includegraphics[width=\columnwidth]
    {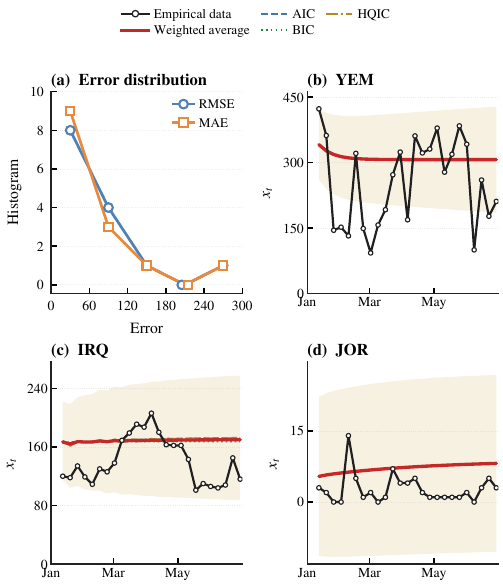}
    \caption{Evaluation results for 2025 obtained from the selected ARIMA parameter sets. (a) Distributions of the prediction errors between the empirical $x_t$ and the weighted-average prediction $\hat{x}_{t,{\rm avg}}$ in Eq.~(\ref{eq:weighted_average}), measured by RMSE and MAE. (b--d) Three representative evaluation results for the original conflict-frequency series $x_t$. For each country, the three dashed lines represent predictions from the ARIMA$(p,d,q)$ models selected by AIC, BIC, and HQIC, and the solid line represents their weighted-average prediction. The shaded area indicates the 95\% confidence interval obtained from the HQIC-selected model, which is used here as a representative prediction interval.}
    \label{fig:forecast}
\end{figure}

\begin{table}[t]
\centering
\caption{RMSE and MAE between the inverse-RMSE weighted-average prediction $\hat{x}_{\mathrm{avg}}$ [Eq.~(\ref{eq:weighted_average})] and the empirical series $x$ for the 2025 evaluation period. Countries are ordered by decreasing RMSE.}
\label{tab:arima_forecast_error}
\setlength{\tabcolsep}{5.0pt}
\begin{tabular}{c|cc}
\hline
Country & RMSE & MAE \\
\hline
LBN & 275.60 & 243.30 \\
SYR & 143.58 & 129.29 \\
YEM & 111.70 & 90.80 \\
PSE & 98.20 & 79.92 \\
IRN & 75.33 & 42.27 \\
ISR & 70.00 & 66.13 \\
IRQ & 41.14 & 35.47 \\
TUR & 41.02 & 30.83 \\
BHR & 9.72 & 8.03 \\
JOR & 5.44 & 5.05 \\
SAU & 0.84 & 0.56 \\
QAT & 0.59 & 0.25 \\
OMN & 0.23 & 0.19 \\
ARE & 0.21 & 0.13 \\
\hline
\end{tabular}
\end{table}

As an additional examination of the linear temporal structure identified from the 2017--2024 training data, we apply the selected ARIMA models to 2025 using the $h$-step-ahead prediction in Eqs.~(\ref{eq:ARIMA_predict})--(\ref{eq:residual}). For $d=0$, the predicted stationary series directly corresponds to the original series $x_t$. For $d=1$, the ARIMA model predicts the first difference $y_t=x_t-x_{t-1}$, and the predictions are transformed back to the original scale by cumulative summation, i.e., $\hat{x}_{T+h|T}=x_T+\sum_{r=1}^{h}\hat{y}_{T+r|T}$.

When different information criteria select different $(p,q)$, they produce different prediction curves. To obtain a representative curve from these parameter sets, we follow the general idea of performance-based forecast combination~\cite{Bates1969Combination}. For each obtained parameter set $i$, the prediction error on the original scale is measured by the root-mean-square-error (RMSE) defined as 
\begin{equation}
\mathrm{RMSE}_i=
\sqrt{
\frac{1}{N^{(c)}}
\sum_{t=T_0}^{T_0+N^{(c)}-1}
\left(x_t-\hat{x}_{t|T,i}\right)^2
},
\label{eq:rmse}
\end{equation}
where $T_0$ denotes the beginning of the evaluation period and $N^{(c)}$ is the number of evaluation points, and the normalized inverse-RMSE weight is
\begin{equation}
w_i=\frac{(\mathrm{RMSE}_i)^{-1}}{\sum_{j\in\mathcal{J}^{(c)}}(\mathrm{RMSE}_j)^{-1}},
\label{eq:weight}
\end{equation}
where $\mathcal{J}^{(c)}$ denotes the set of distinct ARIMA parameter sets for country $c$ that show no significant residual autocorrelation. If different information criteria select the same $(p,d,q)$, the corresponding parameter set is included only once. The weighted-average prediction is then
\begin{equation}
\hat{x}_{t,\mathrm{avg}}=\sum_{j\in\mathcal{J}^{(c)}}w_j\hat{x}_{t|T,j}.
\label{eq:weighted_average}
\end{equation}
Because the weights are obtained from the same prediction period, this weighted curve is an ex-post combination used only as a representative prediction rather than a purely prospective forecast.

The RMSE and mean absolute error (MAE) of $\hat{x}_{t,\mathrm{avg}}$ are listed in Table~\ref{tab:arima_forecast_error}, and their distributions are shown in Fig.~\ref{fig:forecast}(a). Representative results for YEM, IRQ, and JOR are shown in Figs.~\ref{fig:forecast}(b)--\ref{fig:forecast}(d), with RMSE values of 111.7, 41.14, and 5.44, respectively; results for all 14 countries are presented in Fig.~\ref{fig:appendix_arima_all14} in Appendix~\ref{app:all}. Since the weighted-average curve does not have a directly associated prediction interval, we show the 95\% confidence interval of the HQIC-selected ARIMA model as a representative uncertainty range.

The prediction results provide an additional view of how far the linear temporal structure identified in the training data persists into the subsequent period. For several countries, the empirical observations remain broadly consistent with the predicted level, particularly when the RMSE is relatively small, whereas detailed fluctuations and some abrupt changes are not reproduced. Such limitations are expected because ARIMA extrapolates linear temporal dependence inferred from past observations and cannot anticipate changes absent from the training data. Thus, these results are intended as a supplementary examination of the identified linear temporal structure rather than a demonstration of accurate forecasting.

\section{Temporal Persistence of Conflict-Activity Changes}
\label{sec:tau}
\begin{figure}
    \centering
    \includegraphics[width=\columnwidth]{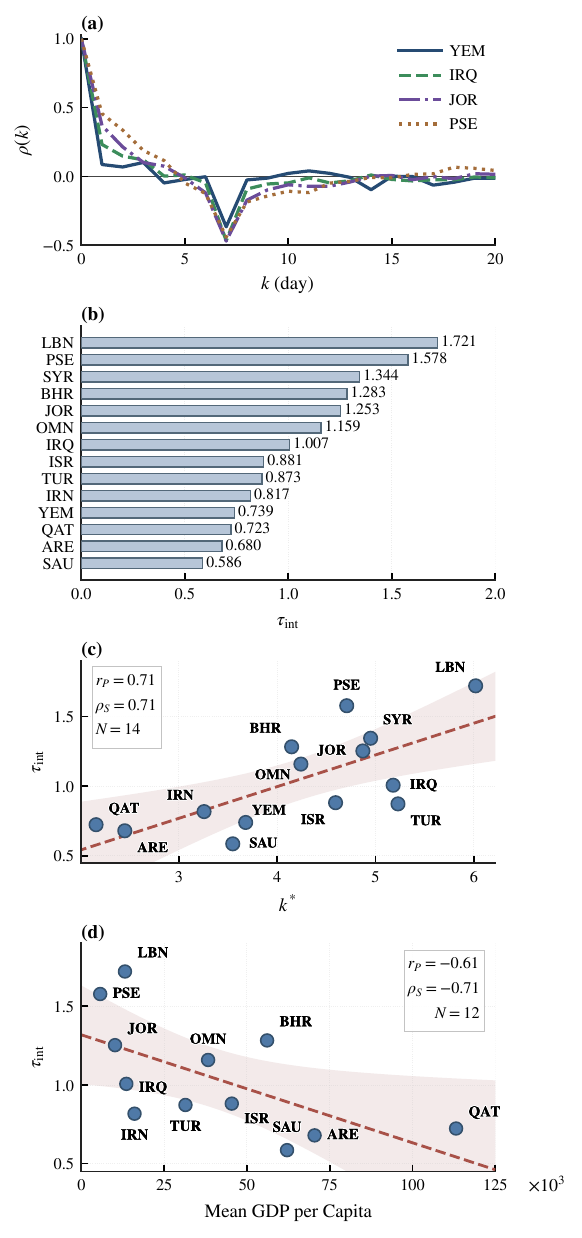}
    \caption{Temporal correlations of local changes in 7-day moving-average conflict activity, $\Delta m_t$. (a) Autocorrelation functions $\rho(k)$ as a function of lag $k$ for YEM, IRQ, JOR, and PSE. (b) Integrated correlation time $\tau_{\rm int}$ [Eq.~(\ref{eq:tauint})] for all 14 countries. (c) Integrated correlation time $\tau_{\rm int}$ versus the first zero-crossing point $k^*$. Significant positive correlations are observed for both Pearson ($r_p=0.71$) and Spearman ($\rho_s=0.71$) measures, with $p<0.05$ for both. (d) Integrated correlation time $\tau_{\rm int}$ versus the mean GDP per capita at PPP over 2017--2024. The GDP comparison includes 12 countries with complete data over the analysis period. Significant negative correlations are observed for both Pearson ($r_p=-0.61$) and Spearman ($\rho_s=-0.71$) measures, with $p<0.05$ for both. The dashed lines and shaded regions represent the linear trends and its 95\% confidence intervals, respectively.
}
    \label{fig:acf_tau}
\end{figure}

\begin{table}[t]
\centering
\caption{Integrated correlation time $\tau_{\rm int}$ (days), first zero-crossing point $k^*$ (days), and mean GDP per capita at PPP ($10^{3}$ constant 2021 international dollars) over 2017--2024. The GDP values are reported for the 12 countries; SYR and YEM are excluded due to a lack of complete data for GDP. They are sorted in terms of $\tau_{\rm int}$.
}
\label{tab:tau}

\setlength{\tabcolsep}{5.0pt}
\begin{tabular}{c|ccc}
\hline
Country
& $\tau_{\mathrm{int}}$
& $k^*$
& Mean GDP \\
\hline
LBN & 1.721 & 6.019 & 13.18  \\
PSE & 1.578 & 4.707 & 5.70   \\
SYR & 1.344 & 4.950 & --     \\
BHR & 1.283 & 4.147 & 56.09  \\
JOR & 1.253 & 4.871 & 10.18  \\
OMN & 1.159 & 4.241 & 38.28  \\
IRQ & 1.007 & 5.179 & 13.59  \\
ISR & 0.881 & 4.594 & 45.43  \\
TUR & 0.873 & 5.229 & 31.44  \\
IRN & 0.817 & 3.256 & 16.07  \\
YEM & 0.739 & 3.680 & --     \\
QAT & 0.723 & 2.158 & 113.16 \\
ARE & 0.680 & 2.450 & 70.43  \\
SAU & 0.586 & 3.548 & 62.11  \\
\hline
\end{tabular}
\end{table}

ARIMA characterizes linear temporal dependence and predictability but does not directly provide a characteristic time scale of temporal persistence. Previous studies have examined temporal dependence in conflict occurrence, fatalities, and sequences of individual events~\cite{Weidmann2010predicting,Fujita2017correlations,Lewis2012Iraq,Kushwaha2023Mesoscale}. Here, we focus instead on how long short-term changes in conflict activity remain correlated. Rather than the persistence of the conflict-frequency level itself, we examine the persistence of increases and decreases, because this captures whether the direction of recent changes in conflict activity continues over short time scales. Using the daily conflict count $n_s^{(c)}$, we construct a 7-day moving average of conflict activity and shift the window in one-day increments, $s=s_0,s_0+1,s_0+2,\ldots$. The local change in conflict activity is defined as the difference between two successive 7-day moving averages,
\begin{equation}
\Delta m_s^{(c)}=\frac{n_{s+4}^{(c)}-n_{s-3}^{(c)}}{7},
\label{eq:delta_mt}
\end{equation}
where the two 7-day windows are separated by one day and overlap over six days.

We calculate the autocorrelation function $\rho(k)$ of $\Delta m_t$ for all countries, with representative cases shown in Fig.~\ref{fig:acf_tau}(a) and the full results in Fig.~\ref{fig:ACF_14countries} in Appendix~\ref{app:all}. In all cases, $\rho(k)$ is positive over the first few lags and subsequently becomes negative. We define the first zero-crossing point $k^*$ by 
\begin{equation}
    \rho(k^*)=0,
    \label{eq:kstar}
\end{equation}
following the common use of the first zero crossing as a practical cutoff for the initial positive-correlation regime in autocorrelation-based time-scale analyses~\cite{Andersson2011,Chi2023}. We estimate $k^*$ by linear interpolation between the two consecutive daily lags across which $\rho(k)$ changes sign. The integrated correlation time is then
\begin{equation}
\tau_{\rm int}=\int_0^{k^*}\rho(k)\,dk.
\label{eq:tauint}
\end{equation}
As shown in Fig.~\ref{fig:acf_tau}(b), $\tau_{\rm int}$ ranges from approximately 0.6 to 1.7 days, indicating that changes in weekly-scale conflict activity remain correlated only over a few days.

The first zero-crossing point $k^*$ and the integrated correlation time $\tau_{\rm int}$ characterize different aspects of short-term temporal persistence. While $k^*$ measures the extent of the positive-correlation regime, $\tau_{\rm int}$ reflects both its extent and the magnitude and decay of $\rho(k)$ within this regime. Therefore, a nearly perfect correlation between the two quantities is not expected, and Fig.~\ref{fig:acf_tau}(c) indeed shows some variation in $\tau_{\rm int}$ for similar values of $k^*$. Nevertheless, they exhibit a clear positive correlation. This relation is natural because $\tau_{\rm int}$ integrates the positive correlation up to $k^*$, so a longer positive-correlation regime generally contributes to a larger $\tau_{\rm int}$.

As an exploratory comparison with a macroscopic socioeconomic indicator, we examine the relation between $\tau_{\rm int}$ and the mean GDP per capita at purchasing-power parity (PPP) over 2017--2024, obtained from the World Bank World Development Indicators~\cite{WorldBankWDI2026}. Due to insufficient data, SYR and YEM are excluded from this analysis, resulting in 12 countries. GDP is used as an external socioeconomic indicator because commonly used political-stability indices already incorporate conflict-related information~\cite{Kaufmann2024WGI}. The resulting $k^*$ and $\tau_{\rm int}$ values, together with GDP, are listed in Table~\ref{tab:tau}. Figure~\ref{fig:acf_tau}(d) shows a significant negative correlation between $\tau_{\rm int}$ and GDP, indicating that changes in conflict activity decorrelate more rapidly in countries with higher GDP. This suggests a possible association between short-term conflict persistence and broader socioeconomic conditions, without implying causality.

\section{Discussion}
\label{sec:discussion}

In this study, we have investigated the linear temporal structure of conflict events in Middle Eastern countries using the ARIMA framework as a simple linear baseline. We have used three information criteria to select the ARIMA parameter sets and found that the linear temporal dependence is largely captured for most countries. Despite the presence of such structure in past observations, predictions beyond the training period reproduce the overall temporal behavior only to some extent, while detailed or abrupt changes remain difficult to predict. We have further examined the characteristic time scales of changes in conflict activity using the first zero-crossing point $k^*$ and the integrated correlation time $\tau_{\rm int}$. The former ranges from approximately 2 to 6 days, while $\tau_{\rm int}$ ranges from approximately 0.6 to 1.7 days. The two quantities are positively correlated, and $\tau_{\rm int}$ further shows a negative association with mean GDP per capita PPP, suggesting a possible relation between short-term conflict persistence and broader socioeconomic conditions.

Unlike more flexible nonlinear prediction approaches, such as reservoir computing and recurrent neural networks, ARIMA is based on a simple linear assumption and therefore has intrinsic limitations in reproducing complex temporal fluctuations. Nevertheless, our results show that a simple linear structure captures a substantial part of the temporal dependence in most cases, despite the apparently complex temporal fluctuations of conflict activity. Identifying such linear structure in seemingly complicated conflict phenomena may provide a useful baseline for developing more refined, system-specific models that incorporate additional nonlinear or external effects.

Despite the rich information provided by ACLED, our analysis has several limitations. First, the present time series represents the frequency of dated ACLED event records rather than the duration of underlying conflict episodes. Thus, the duration of conflicts, despite being an important quantity of interest, cannot be directly characterized from the present data, and the correlation time obtained here should not be interpreted as the duration of individual conflicts. Second, all events are treated equally, although their impacts may differ substantially. Fatality information could be used to weight event magnitude, but it is incomplete for some records. More complete fatality data would allow magnitude-weighted time series and a more quantitative characterization of conflict intensity. Finally, events are assigned to the locations where they occur, so cross-country interactions are not directly represented. Constructing country-level conflict networks would require systematic mapping between organization-level actors and countries, which could be addressed in future work using entity-matching approaches, including large language models.

\section*{Acknowledgement}
This work was supported by the National Research Foundation (NRF) of Korea through Grant Number RS-2024-00341317 (M.J.L.).



\appendix
\setcounter{section}{0}
\renewcommand{\thesection}{\Alph{section}}
\renewcommand{\thesubsection}{\thesection.\arabic{subsection}}

\setcounter{figure}{0}
\renewcommand{\thefigure}{\thesection.\arabic{figure}}

\setcounter{table}{0}
\renewcommand{\thetable}{\thesection.\arabic{table}}

\section{Country code and name}
\label{app:code}
The country code and names are listed in Table~\ref{tab:code}.

\begin{table}[t]
\centering
\caption{Country codes used in this study and their corresponding country names.}
\label{tab:country_codes}
\begin{tabular}{ll}
\hline
Country code & Country name \\
\hline
ARE & United Arab Emirates \\
BHR & Bahrain \\
IRN & Iran \\
IRQ & Iraq \\
ISR & Israel \\
JOR & Jordan \\
KWT & Kuwait \\
LBN & Lebanon \\
OMN & Oman \\
PSE & Palestine \\
QAT & Qatar \\
SAU & Saudi Arabia \\
SYR & Syria \\
TUR & Türkiye \\
YEM & Yemen \\
\hline
\end{tabular}
\label{tab:code}
\end{table}

\section{MODEL SELECTION AND RESIDUAL DIAGNOSTICS}
\label{app:ARIMA_diagnostics}

In this section, we briefly describe some statistical tests.

\subsection{ Information criteria}
\label{app:ic}
For a model with $k$ parameters fitted to $N$ data points, let $\hat{L}$ denote the maximized likelihood. The three information criteria, which are widely used for model selection by balancing goodness of fit and model complexity, are given by
\begin{align}
    \mathrm{AIC} &= 2k - 2\ln\hat{L}, \nonumber \\
    \mathrm{BIC} &= k\ln N - 2\ln\hat{L}, \nonumber \\
    \mathrm{HQIC} &= 2k\ln(\ln N) - 2\ln\hat{L}. \nonumber
    \label{eq:AIC}
\end{align}
All three criteria balance goodness of fit, represented by the likelihood term, against model complexity, represented by the penalty term. AIC imposes a relatively weak penalty on model complexity and thus places more emphasis on goodness of fit, whereas BIC imposes a stronger penalty and tends to favor simpler models. HQIC provides an intermediate balance between AIC and BIC, with a complexity penalty generally stronger than that of AIC but weaker than that of BIC.

\subsection{ Temporal correlation of residuals}
\label{app:ljungbox}
\begin{figure}
\centering \includegraphics[width=\linewidth] {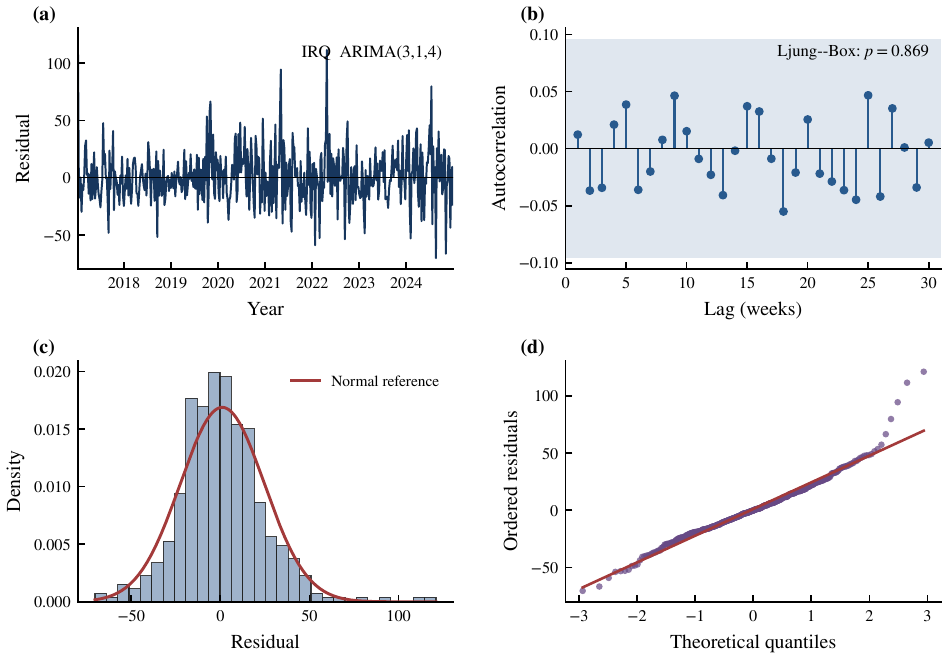} \caption{ Residual diagnostics for the AIC-selected ARIMA$(3,1,4)$ model of IRQ fitted to the 2017--2024 training data. (a) Residual time series. (b) Autocorrelation function of residuals. The Ljung--Box test was evaluated at lag $h=20$ with the model degrees of freedom adjusted by $p+q$. (c) Residual distribution with a normal reference, and (d) Normal Q--Q plot.} 
\label{fig:arima_residual}
\end{figure}

If the fitted ARIMA$(p,d,q)$ adequately captures the linear temporal dependence in the training data, its residuals $\hat{\epsilon}_t$ should show no systematic temporal correlation. This is examined using the Ljung--Box test~\cite{Ljung1978Box}, whose null hypothesis is that the residuals have no significant autocorrelation over the considered range of time lags. As an example, Fig.~\ref{fig:arima_residual}(a) shows the residuals of ARIMA$(3,1,4)$ selected by AIC for IRQ, fitted to the 2017--2024 training data, and Fig.~\ref{fig:arima_residual}(b) shows their autocorrelation. The residual autocorrelations lie within the 95\% confidence interval, and the Ljung--Box test gives $p=0.869$, providing no evidence of remaining temporal correlation. The residual distribution is approximately Gaussian around its central region, but deviations are visible in the tails, as seen in the histogram and Q--Q plot in Figs.~\ref{fig:arima_residual}(c) and~\ref{fig:arima_residual}(d). This deviation does not necessarily indicate model inadequacy, but rather suggests that extreme fluctuations are not fully described by a Gaussian error distribution.

We performed the same residual diagnostics for all countries. Most selected parameter sets pass the Ljung--Box test, while the few exceptions with $p<0.05$ are marked by cross symbols in Table~\ref{tab:arima_models}.

\section{ Results for all countries}
\label{app:all}
\begin{figure*}[p]
    \centering
    \includegraphics[
        width=0.98\textwidth,
        height=0.84\textheight,
        keepaspectratio
    ]{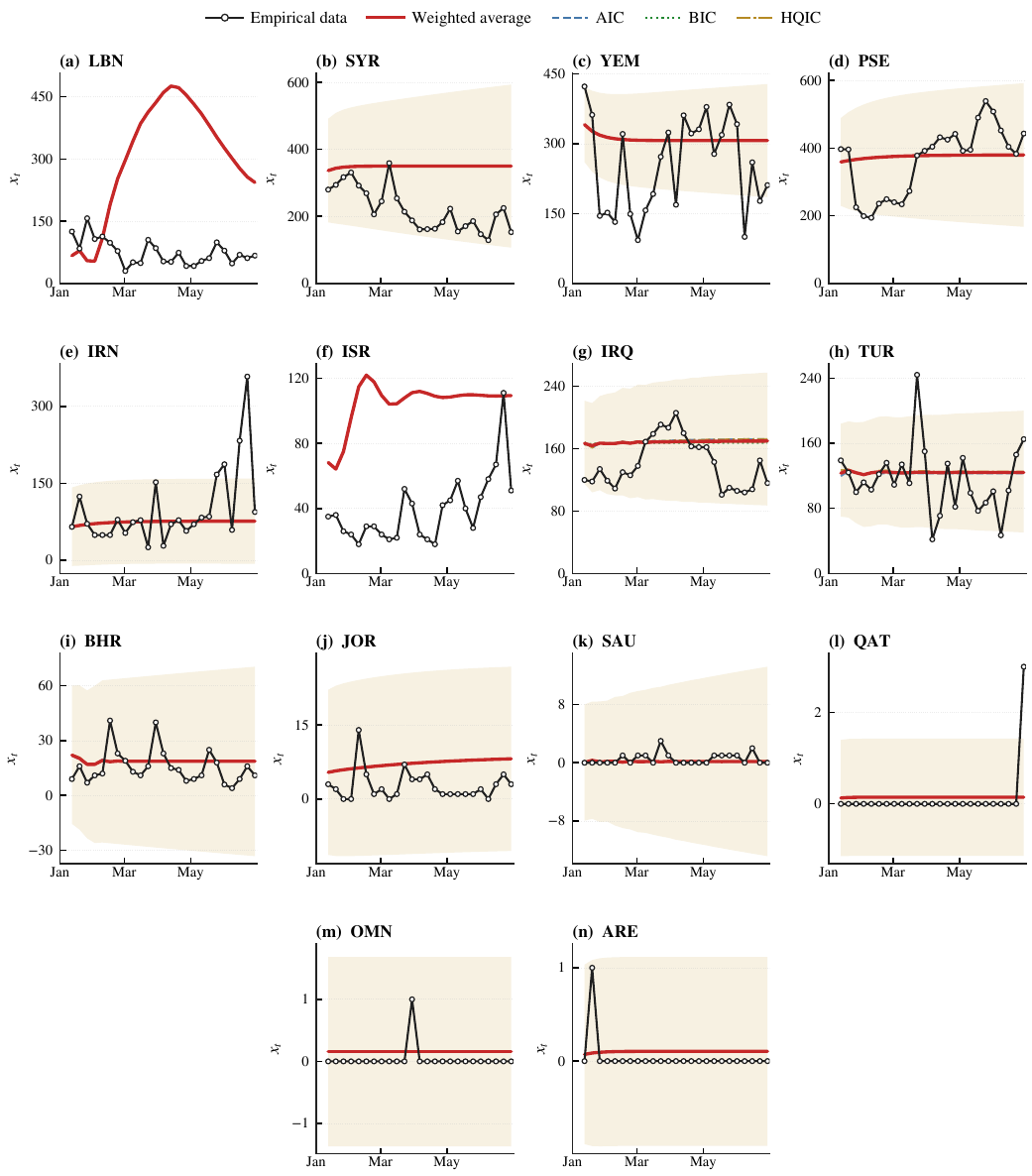}
    \caption{Prediction results for all countries obtained from ARIMA, sorted by RMSE from largest to smallest. Only parameter sets that pass the Ljung--Box test are shown. The shaded areas represent the 95\% confidence intervals from HQIC and are omitted when the HQIC-selected parameter set fails the Ljung--Box test.}
    \label{fig:appendix_arima_all14}
\end{figure*}

\begin{figure*}[!t]
    \centering
    \includegraphics[width=\textwidth]{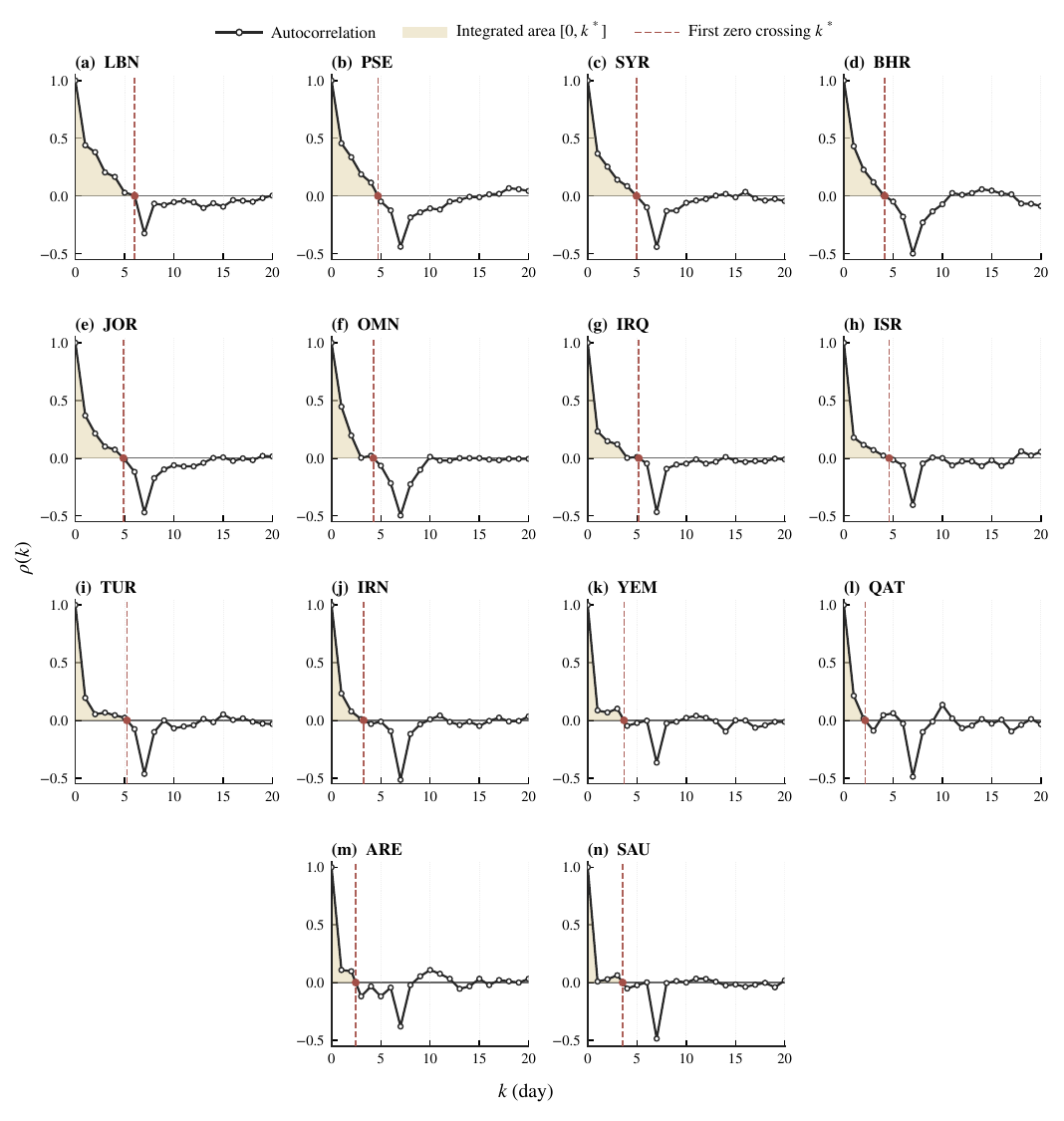}
    \caption{
    Autocorrelation functions $\rho(k)$ of the local changes
    $\Delta m_t$ for all 14 countries.
    The shaded region denotes the piecewise-linear area integrated
    from $k=0$ to the linearly interpolated first zero-crossing point
    $k^*$, corresponding to the integrated correlation time
    $\tau_{\rm int}$ defined in Eq.~(\ref{eq:tauint}).
    The dashed vertical line indicates $k^*$.
    Countries are ordered by decreasing $\tau_{\rm int}$.
    }
    \label{fig:ACF_14countries}
\end{figure*}

In this section, we show all respective curves for all 14 countries. The prediction results by ARIMA$(p, d, q)$ are plotted in Fig.~\ref{fig:appendix_arima_all14}, and the autocorrelation functions are plotted in Fig.~\ref{fig:ACF_14countries}.

\end{document}